\documentclass[a4paper,10pt]{article}
\usepackage{cite}
\usepackage{graphicx}
\usepackage{amssymb}
\usepackage{amsmath}

\begin{document}
\title{Direction dependence of the power spectrum and its effect on the Cosmic Microwave Background Radiation}
\author{Pranati K. Rath$^1$, Tanmay Mudholkar$^1$, Pankaj Jain$^1$,\\ Pavan K. Aluri$^2$, Sukanta 
Panda$^2$}

\maketitle

\begin{center}
{$^{1}$ Department of Physics, Indian Institute of Technology, Kanpur 208016, India \\
$^2$Department of Physics, IISER Bhopal, Bhopal-462 023, India\\}
\end{center}


\begin{abstract}
We study several anisotropic inflationary models and their implications for the observed
violation of statistical isotropy in the CMBR data. In two of these models the anisotropy
decays very quickly during the inflationary phase of expansion. We explicitly show that
these models lead to violation of isotropy only for low $l$ CMBR modes. Our primary aim
is to fit the observed alignment of $l=2,3$ multipoles to the theoretical models. We use
two measures, based on the power tensor, which contains information about the alignment
of each multipole, to quantify the anisotropy in data. One of the measures uses the
dispersion in eigenvalues of the power tensor. We also define another measure which tests
the overall correlation between two different multipoles. We perturbatively compute these
measures of anisotropy and fix the theoretical parameters by making a best fit to $l=2,3$
multipoles. We show that some of the models studied are able to consistently explain
the observed violation of statistical isotropy.
\end{abstract}


\section{Introduction}

There exist many observations which suggest violation of the cosmological principle.
These include dipole anisotropy in radio polarizations \cite{Birch1982, Jain1999},
large scale alignment of optical polarizations \cite{Huts1998}, alignment of $l=2,3$
multipoles of the cosmic microwave background radiation \cite{Costa2004,Copi2004,Abramo2006},
dipole anisotropy in distribution of radio galaxies \cite{Singal2011} and cluster
peculiar velocities \cite{Kashlinsky2010}. Remarkably all of these indicate a preferred
direction towards the Virgo cluster, close to the observed CMB dipole. This coincidence
of diverse axes was first pointed out in \cite{Ralston2004,Schwarz2004}.

There also exit several other indications of violations of isotropy. These include
dipole modulations to CMB including north-south ecliptic power asymmetry
\cite{Eriksen2004,Bernui2008,Erickcek2008,Lew2008,Hansen2009,Hanson2009,Groeneboom2010},
quadrupolar power modulation to CMB \cite{Hanson2009,ACW2007,Groeneboom2008,Bennett2011},
dipole anisotropy in galaxy distribution \cite{Itoh2010}, and parity violation in
the CMB data  \cite{Kim2010,Aluri2012a} and in the handedness of
the spiral galaxies \cite{Longo2011}.
The indications for a preferred direction in diverse data sets have motivated many
theoretical studies of inflationary models which violate statistical isotropy and
homogeneity, giving rise to a direction dependent power spectrum \cite{ACW2007,
Goldwirth1990, Gordon2005, Emir2007, Pontzen2007, Pereira2007, Pullen2007,
Campanelli2009, Donoghue2009, Watanabe2009, KimHC2010, Ma2011, Wang2012}.
Anisotropic inflation is likely to lead to observable effects in the
CMBR \cite{Watanabe2011}.

It is well accepted that the observed Universe is neither isotropic nor homogeneous
at early times. It may evolve into a de Sitter space as it expands in presence of a
positive cosmological constant. This has been explicitly established for Bianchi models,
which represent a homogeneous and anisotropic Universe. An anisotropic Bianchi model
is expected to become isotropic within a time scale of order $\sqrt{\frac{3}{\Lambda}}$
during the early stages of inflation \cite{Wald1983}. It was pointed out in \cite{Aluri2012b},
that even this brief early phase of anisotropic expansion can lead to
violations of statistical isotropy in the current era. 
It was shown that for a wide range of parameters,
modes generated during the anisotropic inflationary period can re-enter the horizon at
recent times and hence affect observations. For some parameter range they may re-enter
even before decoupling of radiation from matter \cite{Aluri2012b}. In the current paper
we shall show that the anisotropic modes arising during the very early phase of inflation
can consistently explain the alignment of CMB quadrupole and octopole. 
We point out that our explanation is 
within the framework of
the inflationary Big Bang cosmology. 

There exist several techniques to study the violation of statistical isotropy in the CMB
data. These include bipolar spherical harmonics \cite{Hajian2003,Hajian2005},
the power tensor \cite{Ralston2004,Samal2008}, multipole vectors \cite{Copi2004} etc.
Here we shall use the power tensor, which is defined by,
\begin{equation}
 A_{ij}(l) = \frac{1}{l(l+1)}\sum_{m,m',m''}\langle{lm|J_{i}|lm'}\rangle\langle{lm''|J_{j}|lm}\rangle a_{lm'}a^{*}_{lm''}
\end{equation}
where $a_{lm}$ are the standard harmonic coefficients of the temperature anisotropy,
$\Delta T (\hat{n})/T$, and $J_{i}$ are the angular momentum operators in spin$-l$ representation.
This tensor is a second rank real symmetric matrix. Its eigenvectors define an invariant
frame for each multipole and the corresponding eigenvalues yield the power along each
of these vectors. For a given multipole, the eigenvector associated with the largest
eigenvalue of the power tensor is called the principal eigenvector (PEV). This defines
the preferred direction for that multipole \cite{Ralston2004,Samal2008}. Statistical
isotropy implies that the two point correlations of the harmonic coefficients obey,
\begin{equation}
\langle{a_{lm}a^*_{l^{\prime}m^{\prime}}}\rangle = \delta_{ll^{'}}\delta_{mm^{'}}C_{l}
\end{equation}
where $C_{l}$ is the power spectrum and the angular brackets denote an ensemble average.
For the power tensor, this implies,
\begin{equation}
 \langle{A_{ij}(l)}\rangle = \frac{C_{l}}{3}\delta_{ij} \, .
\label{eq:Aij}
\end{equation}
 In the case of an anisotropic background metric,
\begin{equation}
\left\langle{a_{lm}a^*_{l^{\prime}m^{\prime}}}\right\rangle = C_{ll'mm'} \, ,
\end{equation} 
i.e. the correlations are not diagonal as in the isotropic case. This will imply that,
in contrast to Eq. \ref{eq:Aij}, the power tensor is not proportional to $\delta_{ij}$.
Furthermore it will also generate correlations in the power tensor for different
multipoles.

The alignment of the CMB quadruple and octopole can be tested in the power tensor
approach by determining the corresponding PEVs \cite{Ralston2004,Samal2008,Aluri2011}.
These are found to be aligned very closely with one another. In the 7 year WMAP data
release, the significance of alignment is found to be better than $4\sigma$ CL in the ILC
map \cite{Aluri2011}. The multipoles, $l=2,3$, by themselves, however, do not give any
indication of violation of statistical isotropy. The hot and cold spots, both for $l=2$
and $l=3$ do lie roughly in a single plane. However the effect is not statistically significant. It is only the alignment of the two multipoles which is significant. 

In our theoretical approach we work in a preferred coordinate system with the preferred
axis aligned along the PEV corresponding to $l=2$. We use two different approaches to
test for alignment. In one approach we determine the PEV for each multipole and then
test for its alignment. In the second approach we follow a blind procedure which
directly tests the overlap of the two power tensors. The overlap is defined as the
trace of the matrix ${\langle{A^{+}(l)A(l')}\rangle}_{ij}$. We may express this
ensemble average as,
\begin{eqnarray}
{\langle{A^{+}(l)A(l')}\rangle}_{ij} &=& \sum_{k}\langle{A_{ki}(l)}^{*}A_{kj}(l')\rangle \nonumber\\
&=&\frac{1}{ N(l)N(l')}\sum_{k}\sum_{m_n=-l}^{l}\sum_{m'_n=-l'}^{l'}{\langle{lm_1|J_k|lm_2}\rangle}^{*}{\langle{lm_3|J_i|lm_1}\rangle}^{*} \nonumber\\
&&{\langle{l'm'_1|J_k|l'm'_2}\rangle}{\langle{l'm'_3|J_j|l'm'_1}\rangle}\langle{a^{*}_{lm_2}a_{lm_3}a_{l'm'_2}a^{*}_{l'm'_3}}\rangle \,.
\end{eqnarray}
where the normalization $N(l)=l(l+1)$.
Here on the right hand side the subscript $n$ takes values $1,2,3$, i.e. the sums are over $m_1,m_2,m_3,m'_1,m'_2,m'_3$. The corresponding statistic is defined as,
\begin{equation}
 S_{ll'} = \frac{3}{C_l C_{l'}}\ Tr{\langle{A^{+}(l)A(l')}\rangle} \, ,
\label{eq:Sllp}
\end{equation}
For the isotropic case, the expectation value on the right hand side 
factorizes and we obtain, 
\begin{equation}
S_{ll'}({\rm isotropic}) = 1\ . 
\end{equation}
Here we shall only be interested in the case $l'=l+1$ and hence use the 
statistic,
\begin{equation}
 S_{l} = \frac{3}{C_l C_{l+1}}\ Tr{\langle{A^{+}(l)A(l+1)}\rangle} \, .
\label{eq:SI}
\end{equation}
The blind
approach has the advantage that it tests the overall correlation of the two multipoles
rather than testing a specific feature. Hence it does not suffer from the criticism of
being an a posteriori statistic \cite{Bennett2011}. As expected, this statistic leads
to a smaller significance of alignment between $l=2,3$.

The power spectrum $P(k)$ for the primordial density perturbation $\delta(k)$ is defined as
\begin{equation}
 \left<\delta(k) \delta^*(k')\right>  =  P(k)\delta^3(k-k')
\end{equation}
where $\delta(k)$ 
is the Fourier amplitude of the density inhomogeneities, 
\begin{equation}
 \frac{\delta \rho(x)}{\rho(x)} = \frac{1}{(2\pi)^{3}}\int \delta(k) e^{-i\vec k\cdot \vec x}d^{3}k \, .
\end{equation}
In the case of a homogeneous and isotropic FRW metric, the fluctuations are statistically
isotropic and the primordial power spectrum $P(k)$ depends only on the magnitude of the
wave vector $\vec k$. However, in the case of anisotropic models, which we shall consider
in this paper, it will also depend on the direction.
The direction dependent contribution to $P(k)$ has been obtained in a 
particular anisotropic model in Ref. \cite{ACW2007}. Here we examine this
model as well as two other models of anisotropic inflation.

\section{Anisotropic Metric}

We consider three different anisotropic metrics. The axis of anisotropy is taken to be the
$z-$axis. The first model may be expressed as,
\begin{equation}
 ds^2 = dt^2 - 2\sqrt{\sigma} dz dt - a^2(t)(dx^2 + dy^2 + dz^2) \quad {\rm Model\ I}
\end{equation}
where $a(t)$ is the scale factor and $\sigma$ is the anisotropic parameter which controls
deviation from statistical isotropy. The second model we consider is 
same as the one
used in \cite{ACW2007}. The corresponding metric is given by,
\begin{equation}
 ds^2 = dt^2 - a^2(t)dx_{\perp}^2 - b^2(t)dz^2 \quad {\rm Model\ II}
\end{equation}
where $dx_\perp^2 = dx^2 + dy^2$. This model is anisotropic throughout inflation.
The third model is a modification of this model such that the anisotropic term becomes
insignificant after one efold. The line elements may be expressed as,
\begin{equation}
 ds^2 = dt^2  - a_{1}^{2}(t)(dx^2 + dy^2 )- a_{2}^2(t)dz^2 \quad {\rm Model\ III}
\end{equation}
where $a_{2}(t) = a_{1}(t)+\sigma$,  $\sigma$ being a constant, independent of time.
We also define, $\bar a(t) = (a_{1}^{2}(t)a_{2}(t))^{\frac{1}{3}}$.

Here we shall present the calculation of the alignment between multipoles for the
case of model I in detail. The calculation for model II and III proceeds along the
same line and for these models we only give the final results. We shall treat the anisotropic
term perturbatively, following the treatment of \cite{ACW2007}. For models I and III this term is expected to be significant
only at very early times.
For model II, this term would contribute during the entire inflationary
period. In all the models, the anisotropy parameter may be fixed by fitting the
data for $l=2,3$ multipoles. The reliability of perturbation theory will depend
on the magnitude of this parameter.

The components of the Einstein tensor for the perturbed metric corresponding to
model I are
\begin{eqnarray}
G_{00} &=& \frac{a(t)[5\sigma+3a^2(t)]{\dot a}^2(t)-2\sigma[\sigma+a^2(t)\ddot a(t)]}{a(t)(\sigma+a^2(t))^2}\\
G_{03} &=& -\sqrt{\sigma}\ \frac{[3\sigma+a^2(t)]{\dot a}^2(t)+2a(t)[\sigma+a^2(t)]\ddot a(t)}{[\sigma+a^2(t)]^2}\\
G_{11} &=& -a^2(t)\ \frac{[3\sigma+a^2(t)]\dot a^2(t)+2a(t)[\sigma+a^2(t)]\ddot a(t)}{[\sigma+a^2(t)]^2} \\
G_{30} &=& G_{03} \\
G_{11} &=& G_{22} = G_{33} \, .
\end{eqnarray}
The perfect fluid energy momentum tensor is defined as,
\begin{equation}
 T_{\mu \nu} = (\rho + P)u_{\mu} u_{\nu} - P g_{\mu\nu} \, .
\end{equation}
This leads to the following off-diagonal terms,
\begin{equation}
 T_{03} = T_{30} = -\rho\sqrt{\sigma} \, .
\end{equation}
The Einstein equations, to zeroth order in the anisotropic parameter $\sqrt{\sigma}$,
are  given by,
\begin{eqnarray}
\left(\frac{\dot a}{a}\right)^2 + 2\frac{\ddot a}{a} &=& -8\pi G P \\
3\left(\frac{\dot a}{a}\right)^2 &=&  8\pi G \rho
\end{eqnarray}
where the dot refers to the derivative with respect to the cosmic time $t$.
Hence with $P = -\rho$ we get standard inflationary solution,
\begin{equation}
 a(t) = a_{I} e^{Ht}
\end{equation}
where $H$ is the Hubble constant during inflation.

\section{Quantization of field and power spectrum}
In this section we discuss the quantization of the massless scalar field \cite{Dodelson}
and obtain the power spectrum perturbatively to first order in the anisotropy
parameter, $\sqrt{\sigma}$. The action may be written as,
\begin{equation}
S_{\phi} = \int d^{4}x\,\mathcal{L}
\end{equation}
where the Lagrangian density,
\begin{equation}
 \mathcal{L} = \frac{1}{2}{\sqrt{-g}}g^{\mu\nu}\partial_{\mu}\phi\partial_{\nu}\phi \, .
\end{equation}
The scalar field $\phi(x,t)$ is quantized in the standard manner by expanding in Fourier
modes,
\begin{equation}
\phi_I(x,t) = \int\frac{d^{3}k}{(2\pi)^3}\left(e^{i \vec k \cdot \vec x} \phi_k(t) a_{k} 
+ e^{-i \vec k \cdot \vec x} \phi^*_k(t) a^{\dagger}_{k}\right)\, .
\label{eq:FourierDec}
\end{equation}
The creation and annihilation operators $a^{+}_{k}$ and $a_{k}$ satisfy the commutation
relation $[a_{k},a_{k'}^{+}] = {(2\pi)}^3\delta(k-k')$. The Euler-Lagrange equation of
motion,
\begin{equation}
 \frac{1}{\sqrt{-g}}\partial_{\nu}(g^{\mu\nu}\sqrt{-g}\partial_{\mu}\phi)=0 \, ,
\end{equation}
for the isotropic FRW metric reduces to, 
\begin{equation}
\ddot \phi+3H\dot \phi-\frac{{\bigtriangledown}^2}{a^2}\phi=0 \, .
\end{equation}
In conformal time, this can be written as,
\begin{equation}
 v_{k}''+\left(k^2-\frac{a''}{a}\right)v_{k} = 0 \, ,
\end{equation}
where, 
$v_{k} = a \phi_{k}$ 
and $'=\frac{d}{d\eta}$ denotes the derivative with respect to the conformal time.
The solution to the unperturbed equations, $\phi_{k}^{0}(\eta)$, is given by,
\begin{equation}
\phi^0_{k}(\eta) = \frac{H}{\sqrt{2k}}\left(\frac{i}{k}-\eta\right)\exp({-ik\eta}) \, .
\end{equation}
The Hamiltonian in interaction-picture is found to be
\begin{equation}
H_{I} =\int{d^{3}x} \, {\sqrt{\sigma}}
\left(\frac{d\phi_{I}(x,\eta)}{d\eta} \frac{d\phi_{I}(x,\eta)}{dz} \right ) \, .
\end{equation}
The two point correlations, to first order, are given by, 
\cite{ACW2007,Weinberg2005}, 
\begin{eqnarray}
\left\langle{\phi(x_1,t)\phi(x_2,t)}\right\rangle &\equiv& \left\langle{\phi_I(x_1,t)\phi_I(x_2,t)}\right\rangle + \nonumber \\
        && {{i}\int_{0}^{t}} dt'\left\langle{\left[H_{I}(t'),\phi_I(x_1,t)\phi_I(x_2,t)\right]}\right\rangle \, .
\end{eqnarray}
This leads to,
\begin{equation}
\left\langle{\phi(x_1,t)\phi(x_2,t)}\right\rangle =
\int\frac{d^{3}k}{({2\pi})^{3}}\exp\left[{i\vec k\cdot(\vec x_1-\vec x_2)}\right]
\left[P_{iso}(k)+ k_z\Delta P(k)\right] 
\end{equation}
where 
\begin{equation}
 P_{iso}(k) \simeq |\phi_{k}^{(0)}(\eta)|^{2}
\end{equation}
and
\begin{equation}
\Delta P(k)=-2{\sqrt{\sigma}}\int_{\frac{-1}{H a_{I}}}^{\eta} a(\eta^{\prime})d{\eta^{\prime}}
\left[ \frac{d\phi_{k}^{0}(\eta^{\prime})}{d\eta^{\prime}}{\phi_{k}^{0}}^{*}(\eta){\phi_{k}^{0}}(\eta^{\prime}){\phi_{k}^{0}}^{*}(\eta)+ c.c\right] \, .
\end{equation} 
Here we have used, 
\begin{equation}
\eta = \int\frac{dt}{a(t)} = -\frac{1}{H a_{I}}e^{-H t} \, .
\end{equation}
The complete expression for $\Delta P(k)$ is given by, 
\begin{eqnarray}
\Delta P(k) &=&P_{iso}(k)\left(\frac{\sqrt{\sigma}}{k}\right)\Bigg\lbrace H\eta(2+k^2{\eta}^2)\nonumber\\
&& +\frac{1}{a_{I}}\left(-1-3a_{I}H\eta+k^2{\eta}^2\right)
\cos\left[2k\left(\frac{1}{a_{I}H}+\eta\right)\right]\nonumber\\
&& -\frac{1}{2a_{I}k}\left[4k^{2}\eta+3a_{I}H(-1+k^{2}\eta^{2})\right]
\sin\left[2k\left(\frac{1}{a_{I}H}+\eta\right)\right]\Bigg\rbrace \, .
\end{eqnarray}
Here we assume that the universe was anisotropic at early times. The perturbations
which are generated during the early anisotropic phase may re-enter the horizon
during matter or radiation dominated phases. These modes crossed the horizon very
early during inflation. This implies $k \sim a_{I}H$. Taking the limit $|{k\eta}|<<1$,
we find that,
\begin{equation}
 P_{iso}(k) \simeq \frac{H^{2}}{2k^{3}} \, ,
\end{equation}
\begin{equation}
\Delta P(k) = -\left(\frac{\sqrt{\sigma}}{a_{I}}\right)
\left(\frac{H^2}{2k^4}\right)\left [\cos \left(\frac{2k}{a_{I}H}\right)-
\left(\frac{3Ha_{I}}{2k}\right)\sin \left(\frac{2k}{a_{I}H}\right)\right] \, .
\end{equation}
Finally, the modified direction dependent power spectrum is given by,
\begin{equation}
P'(k) = P_{iso}(k)[1+(\hat k \cdot \hat z)g(k)] \quad {\rm Model\ I}
\end{equation}
where $\hat k$ is the unit vector along the direction of $k$, 
$\hat z$ is the unit
vector along the preferred direction and the function $g(k)$ is defined as,
\begin{equation}
g(k) = -\frac{\sqrt{\sigma}}{a_{I}}\left [\cos \left(\frac{2k}{a_{I}H}\right )-
\left(\frac{3Ha_{I}}{2k}\right)\sin \left(\frac{2k}{a_{I}H}\right)\right ] \, .
\label{eq:gk}
\end{equation}

The corresponding results in model II have already been obtained in Ref. \cite{ACW2007}.
The power spectrum can be expressed as,
\begin{equation}
 P'(k) = P_{iso}(k)\left[1+g(k)(\hat k \cdot \hat n)^2\right] \quad {\rm Model \ II}
\end{equation}
where $\hat n = (0,0,1)$ in the notation of Ref. \cite{ACW2007}. The function,
$g(k)$, at the end of inflation (time $t_{*}$) is given by, 
\begin{equation}
 g(k) = \frac{9}{2}\epsilon_{H} \log\left[\frac{q(t_{*})}{\bar H}\right], 
\end{equation}
 $\bar H = \frac{1}{3}(2H_a+H_b)$ and the physical wavelength $q(t_{*})  =  \frac{k}{\bar a(t_{*})}$.
The anisotropic parameter $\epsilon_{H}$ is defined as
\begin{equation}
\epsilon_{H} = \frac{2}{3}\left (\frac{H_{b} - H_{a}}{\bar H}\right ) \, .
\end{equation}
Here we have neglected the perturbative correction to the 
isotropic power spectrum.

In the case of model III, the Hamiltonian in interaction picture is 
\begin{eqnarray}
 H_{I}(t) = -\int d^{3}x \frac{1}{2} \left\{ \left( \bar{a}(t)-a_{2}(t) \right) \left[ \left( \frac{d\phi}{dx} \right)^2
            + \left( \frac{d\phi}{dy} \right)^2 \right] \right. \nonumber \\
            + \left. \left( \bar{a}(t) - \frac{a_1^2(t)}{a_2(t)} \right) \left( \frac{d\phi}{dz} \right)^{2} \right\} \, .
\end{eqnarray}
The two point correlation function is found to be,
\begin{equation}
\left\langle{\phi(x_1,t)\phi(x_2,t)}\right\rangle =
\int\frac{d^{3}k}{({2\pi})^{3}}\exp\left[{i\vec k\cdot(\vec x_1-\vec x_2)}\right]
\left[P'_{iso}(k)+ k_{z}^{2}\Delta P(k)\right] 
\end{equation}
where
\begin{equation}
\Delta P(k) \simeq  -2\sigma \int_{\frac{-1}{\bar H a_{I}}}^{\eta} a(\eta^{\prime})d{\eta^{\prime}}
\left [2 Im \left ( {\phi_{k}^{0}(\eta^{\prime})}^2{\phi_{k}^{0}}^{*}(\eta)^2\right ) \right ] \, .
\end{equation} 
The modified power spectrum is given by, 
\begin{equation}
 P'(k) = P'_{iso}(k)\left [1+(\hat k\cdot \hat z)^2 g(k)\right ]\ ,
\end{equation}
where,
\begin{equation}
 g(k) = -\frac{\sigma}{a_Ik} \left [{k\cos\left ( \frac{2k}{a_I\bar H}\right )-\frac {5}{2}a_I\bar H \sin \left (\frac{2k}{a_I\bar H}\right )+2a_I\bar H 
\mathrm{Si}\left(\frac{2k}{a_I\bar H}\right)}\right ]\
\end{equation}
with
\begin{equation}
\mathrm{Si}\left(x\right) \equiv \int_0^x\!\mathrm{d}x^{\prime}\, \frac{\sin{x^{\prime}}}{x^{\prime}} \, .
\end{equation}
Here the isotropic power spectrum, $P'_{iso}(k)$, includes a perturbative 
correction to $P_{iso}(k)$. It is given by,
\begin{equation}
P'_{iso}(k) = P_{iso}(k)\left[1-\frac{g(k)}{3}\right]
\end{equation}

\section{Effect of the power spectrum on the Cosmic Microwave Background}
The temperature perturbation
can be decomposed in terms of spherical
harmonics, $Y_{lm}(\hat p)$, as follows, 
\begin{equation}
 \frac{\delta T}{T}(\vec x,\hat p,\eta) = \sum_{l=1}^{\infty}\sum_{m=-l}^{l}a_{lm}Y_{lm}(\hat p)\ .
\end{equation}
This temperature anisotropy $\frac{\delta T}{T}$ can be related to the primordial
density fluctuations as :
\begin{equation}
 \frac{\delta T}{T}(\hat p)= \int dk\sum_{l}\frac{2l+1}{4\pi}(-i)^{l}P_{l}(\hat k\cdot \hat p)\delta(k)\Theta_{l}(k)
\end{equation}
where $P_{l}$ is the Legendre polynomial of order $l$ and $\Theta_{l}(k)$ is the transfer
function which correlates initial fluctuations to the observed temperature anisotropies.
 The expansion coefficients $a_{lm}$ can be calculated using,
\begin{equation}
 a_{lm} = \int d\Omega Y_{lm}^{*}(\hat p)\frac{\delta T}{T}(\vec x,\hat p,\eta) \, .
\end{equation}
We now compute the two point correlation function of $a_{lm}$'s using the directional
dependent power spectrum. We have, 
\begin{equation}
\langle{a_{lm}a^{*}_{l'm'}}\rangle = {\langle{a_{lm}a^{*}_{l'm'}}\rangle}_{iso}+{\langle{a_{lm}a^{*}_{l'm'}}\rangle}_{aniso} 
\end{equation}
where the first term corresponds to the isotropic case.
It is given by
\begin{equation}
 {\langle{a_{lm}a^{*}_{l'm'}}\rangle}_{iso} = \delta_{ll'}\delta_{mm'}\int_{0}^{\infty}k^{2}dkP_{iso}(k){\Theta^{2}_{l}(k)} \, .
\end{equation}
The second term which contains departures from statistical isotropy is given 
by, 
\begin{equation}
{\langle{a_{lm}a^{*}_{l'm'}}\rangle}_{aniso} = (-i)^{l-l'}\xi_{lm;l'm'}\int_{0}^{\infty}k^{2}dkP_{iso}(k)g(k){\Theta_{l}(k)}{\Theta_{l'}(k)}
\end{equation}
where we have used the spherical components of the unit vector $n$ as
\begin{equation}
 n_{+} = -(\frac{n_{x}-in_{y}}{\sqrt 2}),n_{-} = (\frac{n_{x}+in_{y}}{\sqrt 2}),n_0 = n_{z}\ .
\end{equation}
 The constant $\xi_{lm;l'm'}$,  which encodes the correlation between different
multipoles, is defined as
\begin{eqnarray}
 \xi_{lm;l'm'} &=& \sqrt{\frac{4\pi}{3}}\int d{\Omega_{k}}(Y_{lm}(\hat k))^{*}Y_{l'm'}(\hat k) \nonumber\\
&&\times \left (n_{+}Y_{1}^{1}(\hat k)+n_{-}Y_{1}^{-1}(\hat k)+n_{0}Y_{1}^{0}(\hat k)\right )\nonumber\\
&=& n_{+}\xi^{+}_{lm;l'm'}+n_{-}\xi^{-}_{lm;l'm'}+n_{0}\xi^{0}_{lm;l'm'} \, ,
\end{eqnarray}
where,
\begin{eqnarray}
 \xi^{+}_{lm;l'm'}&=&\sqrt{\frac{1}{2}}\delta_{m',m-1}\Bigg[-\sqrt{\frac{(l-m+1)(l-m+2)}{{(2l+1)}{(2l+3)}}}\delta_{l',l+1} \nonumber\\
&&+\sqrt{\frac{(l+m-1)(l+m)}{{(2l+1)}{(2l-1)}}}\delta_{l',l-1}\Bigg] \\
 \xi^{-}_{lm;l'm'}&=&\sqrt{\frac{1}{2}}\delta_{m',m+1}\Bigg[-\sqrt{\frac{(l+m+1)(l+m)}{{(2l+1)}{(2l+3)}}}\delta_{l',l+1} \nonumber\\
&&+\sqrt{\frac{(l-m-1)(l-m)}{{(2l+1)}{(2l-1)}}}\delta_{l',l-1}\Bigg] \\
 \xi^{0}_{lm;l'm'}&=&\delta_{m',m}\Bigg[\sqrt{\frac{(l-m+1)(l+m+1)}{{(2l+1)}{(2l+3)}}}\delta_{l',l+1} \nonumber\\
&&+\sqrt{\frac{(l-m)(l+m)}{{(2l+1)}{(2l-1)}}}\delta_{l',l-1}\Bigg] \, .
\end{eqnarray}
Similarly, using $a_{lm} = (-1)^{m}a^{*}_{l-m}$, we obtain,
\begin{equation}
 \langle{a_{lm}a_{l'm'}}\rangle_{iso} =  (-1)^{m'}\delta_{l,l'}\delta_{m,-m'}\int_{0}^{\infty}k^{2}dkP_{iso}(k){\Theta^{2}_{l}(k)}
\end{equation}
and
\begin{equation}
 \langle{a_{lm}a_{l'm'}}\rangle_{aniso} =  (-1)^{m'}(-i)^{l-l'}\bar\xi_{lm;l'-m'}\int_{0}^{\infty}k^{2}dkP_{iso}(k)g(k){\Theta_{l}(k)}{\Theta_{l'}(k)} \, .
\end{equation}
We express $\bar\xi_{lm;l'-m'}$ as,
\begin{equation}
 \bar\xi_{lm;l'-m'} 
= n_{+}\bar\xi^{+}_{lm;l'-m'}+n_{-}\bar\xi^{-}_{lm;l'-m'}+n_{0}\bar\xi^{0}_{lm;l'-m'} \, ,
\end{equation}
where,
\begin{eqnarray}
 \bar\xi^{+}_{lm;l'-m'}&=&\sqrt{\frac{1}{2}}\delta_{m',-(m-1)}\Bigg[\sqrt{\frac{(l-m+1)(l-m+2)}{{(2l+1)}{(2l+3)}}}\delta_{l',l+1} \nonumber\\
&&+\sqrt{\frac{(l+m-1)(l+m)}{{(2l+1)}{(2l-1)}}}\delta_{l',l-1}\Bigg] \\
 \bar\xi^{-}_{lm;l'-m'}&=&\sqrt{\frac{1}{2}}\delta_{m',-(m+1)}\Bigg[\sqrt{\frac{(l+m+1)(l+m+2)}{{(2l+1)}{(2l+3)}}}\delta_{l',l+1} \nonumber\\
&&+\sqrt{\frac{(l-m-1)(l-m)}{{(2l+1)}{(2l-1)}}}\delta_{l',l-1}\Bigg] \\
 \bar\xi^{0}_{lm;l'-m'}&=&\delta_{m',-m}\Bigg[\sqrt{\frac{(l-m+1)(l+m+1)}{{(2l+1)}{(2l+3)}}}\delta_{l',l+1} \nonumber\\
&&+\sqrt{\frac{(l-m)(l+m)}{{(2l+1)}{(2l-1)}}}\delta_{l',l-1}\Bigg] \, .
\end{eqnarray}
Computing the trace of ${\langle{A^{+}(l)A(l+1)}\rangle}$, we find
\begin{eqnarray}
Tr{\langle{A^{+}(l)A(l+1)}\rangle}&=&\frac{1}{4l(l+1)^{2}(l+2)}\sum_{m_{i}=-l}^{l}\Bigg[d^{+}(l,m_1)d^{-}(l+1,m_5)\delta_{m_2,m_1+1}\delta_{m_4,m_5-1}\nonumber\\
&&+d^{-}(l,m_1)d^{+}(l+1,m_5)\delta_{m_2,m_1-1}\delta_{m_4,m_5+1}+2m_1m_5\delta_{m_1,m_2}\delta_{m_4,m_5}\Bigg]\nonumber\\
&&\times \Bigg[d^{+}(l,m_3)d^{-}(l+1,m_4)\delta_{m_1,m_{3}+1}\delta_{m_6,m_{4}-1}+d^{-}(l,m_3)d^{+}(l+1,m_4)\nonumber\\
&&\delta_{m1,m_3-1}\delta_{m6,m_4+1}+2m_{3}m_{4}\delta_{m_1,m_3}\delta_{m_6,m_4}\Bigg] \times \nonumber\\ 
&&\Bigg[\delta_{m_2,m_3}\delta_{m_5,m_6}\int_{0}^{\infty}k^2dkP_{iso}(k){\Theta^{2}_{l}(k)}\int_{0}^{\infty}k^{2}dkP_{iso}(k){\Theta^{2}_{l+1}(k)}\nonumber\\
&&+\left (\bar\xi_{lm_2;(l+1)m_6}\bar\xi^{*}_{lm_3;(l+1)m_5}+\xi_{lm_2;(l+1)m_5}\xi^{*}_{lm_3;(l+1)m_6}\right ) \times \nonumber\\
&&\left (\int_{0}^{\infty}k^{2}dkP_{iso}(k)g(k)\Theta_{l}(k)\Theta_{l+1}(k)\right )^{2}\Bigg]\
\label{eq:TrAl1Al}
\end{eqnarray}
where we defined $d^{+}(l,m) = \sqrt{(l-m)(l+m+1)}$ and $d^{-}(l,m) = \sqrt{(l+m)(l-m+1)}$.
Note that the sum in the above equation is over $m_i$, $i=1,2,...,6$.
Here we take the transfer function for large scales as
$\Theta_{l}(k) = \frac{1}{3}J_{l}[k(\eta_{0}-\eta_{d})]$, where $\eta_0$ is
the conformal time today and $\eta_d$ is the conformal time at last scattering.
Since $\eta_d\ll \eta_0 $ we take $J_{l}[k(\eta_{0}-\eta_{d})] \approx J_{l}[k\eta_0]$.
It is convenient to define, $x=k\eta_0$. In terms of $x$ the function $g(k)$ in
Eq. \ref{eq:gk} may be expressed as,
\begin{equation}
g(x) =-\frac{\sqrt{\sigma}}{a_{I}} \left[ \cos\left( \frac{2x}{\eta_{0}a_{I}H}\right ) - 
\frac{3H\eta_{0}a_{I}}{2x}\sin\left (\frac{2x}{\eta_{0}a_{I}H}\right )\right ].
\label{eq:gx}
\end{equation}

The integral in Eq. \ref{eq:TrAl1Al} involves an oscillatory functions of $x$.
It is clear that for large $l$ the integral will get contributions from
relatively large values of $k$. For sufficiently large $k$ the integral will
be damped due to oscillations in $g(x)$. Hence the correlations
between different $l$ are expected to be small for large $l$.

\subsection{Models II and III}
For models II and III the constant $\xi_{lm;l'm'}$ is the same as in \cite{ACW2007}
and is given by
\begin{eqnarray}
 \xi_{lm;l'm'} &=& \frac{4\pi}{3}\int d{\Omega_{k}}(Y_{lm}(\hat k))^{*}Y_{l'm'}(\hat k) \nonumber\\
&&\times \left (n_{+}Y_{1}^{1}(\hat k)+n_{-}Y_{1}^{-1}(\hat k)+n_{0}Y_{1}^{0}(\hat k)\right )^{2}\nonumber\\
&=& n^{2}_{+}\xi^{++}_{lm;l'm'}+n^{2}_{-}\xi^{--}_{lm;l'm'}+2n_{+}n_{-}\xi^{+-}_{lm;l'm'}+2n_{+}n_{0}\xi^{+0}_{lm;l'm'}\nonumber\\
&&+2n_{-}n_{0}\xi^{-0}_{lm;l'm'}+n^{2}_{0}\xi^{00}_{lm;l'm'} \, .
\end{eqnarray}
For $l = l'$ the contributing terms of these coefficients are
\begin{eqnarray}
 \xi^{--}_{lm;l'm'} &=& - \delta_{m',m+2} \, \frac{\sqrt{(l^2-(m+1)^2)(l+m+2)(l-m)}}{(2l+3)(2l-1)} \\
 \xi^{--}_{lm;l'm'} &=& \xi^{++}_{l'm';lm} \\
 \xi^{+-}_{lm;l'm'} &=& - \delta_{m',m} \, \frac{(l^2+m^2+l-1)}{(2l-1)(2l+3)} \\
 \xi^{-0}_{lm;l'm'} &=& -\frac{1}{\sqrt{2}} \, \delta_{m',m+1} \, \frac{(2m+1)\sqrt{(l+m+1)(l-m)}}{(2l-1)(2l+3)} \\
 \xi^{+0}_{lm;l'm'} &=& -\xi^{-0}_{l'm';lm} \\
 \xi^{00}_{lm;l'm'} &=& \delta_{m,m'} \, \frac{(2l^{2}+2l-2m^{2}-1)}{(2l-1)(2l+3)}
\end{eqnarray}
Following \cite{ACW2007} we set, 
$|\log\left(\frac{q(t_{*})}{\bar H}\right)| \simeq 60$.
Hence, $g(k) = g_{*} = \frac{9}{2}\epsilon_{H}\times 60$.
The ensemble average of the power tensor is given by,
\begin{eqnarray}
 \langle A_{ij}(l)\rangle &=&\frac{1}{l(l+1)}\sum_{mm'm''}\langle{lm|J_{i}|lm'}\rangle\langle{lm''|J_{j}|lm}\rangle \nonumber\\
&& \times \Bigg [\langle{a_{lm'}a^{*}_{lm''}}\rangle_{iso}+\langle{a_{lm'}a^{*}_{lm''}}\rangle_{aniso} \Bigg ]\
\label{eq:meanAij}
\end{eqnarray}
In this model we find that $\langle A_{ij}(l)\rangle \neq \frac{C_{l}}{3}\delta_{ij}$.
In the preferred coordinate system we obtain a diagonal matrix with the component
corresponding to the preferred direction different from the other two.

\section{Results}
We first give the numerical results for model I. In this case we use the statistic
$S_l$ (Eq. \ref{eq:SI}), in order to test the correlation between different multipoles.
As explained in the introduction, this tests the overall correlation of the two
multipoles rather than testing a particular aspect, such as the alignment of their
principal eigenvectors. For an isotropic model, this statistic takes the
value unity. We have explicitly verified this by using
randomly generated CMB data. Furthermore the anisotropic term gives zero contribution
to $C_l$ in this model. The value of this statistic for the observed CMB data is given
in the second column in Table 1. Here the results are presented for seven year
WMAP data. We have checked that the nine year data also gives nearly
identical results. The fourth column in the table gives the significance
or P-value, i.e. the probability that the correlation seen in data arises as a random
fluctuation in a statistically isotropic sample. This is computed by comparing the data
statistic with 4000 random realizations of CMBR based on the $\Lambda$CDM model. We find
that this statistic also indicates a significant correlation between $l=2$ and $l=3$
multipoles but doesn't lead to significant result for other multipoles. This agrees
with earlier results which show a very strong alignment of the principal eigenvectors
for $l=2,3$ \cite{Aluri2011} and a relatively mild effect for remaining multipoles
\cite{Samal2008}.

The theoretically computed values of $S_l$ for model I are given in the third column
of Table 1. Here we have taken the anisotropy parameter $\sqrt{\sigma'}= \frac{\sqrt{\sigma}}{a_{I}}= 1.73$
in order to fit the observed $S_l$ for $l=2,3$. We also set $\frac{2}{\eta_{0}a_{I}H} = \frac{1}{2}$.
If we choose a smaller value for this parameter, we shall obtain larger correlations
for larger values of $l$. This choice of parameters fits the observations, as shown in
Table 1. Our theoretical results match the data for $(l,l+1) \ge (3,4)$ in the sense that
neither the observed nor the theoretical values indicate significant correlation. We do
not expect to obtain an exact match since the observed values are statistical in nature.
Assuming that statistical isotropy is valid for $l \ge 4$, we expect that each of these
multipoles would be aligned randomly and hence their correlations are expected to be small
and take random values.  

\begin{table}[ht]
  \centering 
  \begin{tabular}{cccc}
\hline
\hline
(l,l+1) &$S_{l}(Data-value)$&$Theoretical-value$&$P-value$ \\
\hline
\hline
(2,3)&$1.487$&$1.487$&$0.005$\\
(3,4)&$0.696$&$1.279$&$0.969$\\
(4,5)&$1.030$&$1.127$&$0.416$\\
(5,6)&$1.100$&$1.039$&$0.198$\\
(6,7)&$1.139$&$1.004$&$0.102$\\
(7,8)&$1.055$&$1.002$&$0.270$\\
(8,9)&$1.007$&$1.013$&$0.458$\\
(9,10)&$1.050$&$1.021$&$0.227$\\
(10,11)&$0.989$&$1.019$&$0.568$\\
(11,12)&$0.934$&$1.001$&$0.849$\\
(12,13)&$0.886$&$1.003$&$0.968$\\
\hline
\end{tabular}
\caption{Comparison of the observed and theoretical results for model I.
         In column 2 and 3 we show the observed and theoretical values of
         the statistic $S_l$. The significance or P-value of the observed
         statistic is shown in column 4.}
\end{table}

We point out that the contribution due to the anisotropic term is relatively small
compared to unity, obtained with an isotropic model. Hence we expect perturbation
theory to be qualitatively reliable. The maximum contribution is found to be $0.487$
for the case of $l=(2,3)$. We expect higher order corrections of the order of
$(0.487)^2$, roughly about $25\%$.

We next describe the results corresponding to model II. In this case the metric
is anisotropic during the entire inflationary period. Hence we expect that all
the modes, independent of the values of $k$, would violate isotropy and lead
to alignment among CMB multipoles of arbitrary $l$ values. In this case we
compute the principal eigenvectors of the power tensor of each multipole.
For each $l$ we expect two equal eigenvalues. The third eigenvalue is found
to be larger than these two eigenvalues for negative values of the anisotropy
parameter $\epsilon_H$. Hence, in this case, the principal eigenvector would
align with the preferred ($z$) axis. We fix the value of $\epsilon_H$ by demanding
that the theoretical and observed values for dispersion in the eigenvalues match
with one another. A useful measure for dispersion is the power entropy \cite{Samal2008}.
Let $\lambda_a$, $a=1,2,3$ represent the three eigenvalues of the power tensor.
As discussed in \cite{Samal2008}, 
these eigenvalues are positive. We define the normalized eigenvalues
\begin{equation}
\tilde{\lambda}_a = \frac{\lambda_a}{\sum_i \lambda_i}
\end{equation}
The power entropy, $S_P(l)$, may be expressed as,
\begin{equation}
S_P(l) = -\sum_a \tilde{\lambda}_a \log \tilde{\lambda}_a
\end{equation}
$S_P(l)$ can take values in the range $0$ to $\log 3$, with $0$ being the value
for maximal dispersion and, hence, anisotropy. In the CMB data sample, the value
of $S_P(l)$ for any $l$ deviates from the perfectly isotropic limit of $\log 3$
due to cosmic variance. The cosmic variance becomes small for large $l$. Hence,
as expected, the observed value of $S_P(l)$ approaches $\log 3$ for large $l$
\cite{Samal2008}. We compute the power entropy expected for an isotropic sample
by using 4000 randomly generated CMB data samples. Let $S_P(l,data)$ and $S_P(l,mean)$
denote, respectively, the power entropy of the observed CMB data and the mean of
random samples. We attribute the difference $\Delta S_P(l)= S_P(l,data) - S_P(l,mean)$
to the contribution due to the anisotropic term. The theoretical estimate of
$\Delta S_P(l)$ is, therefore, equal to $\Delta S_{P,th}(l)=S_{P,th}-\log 3$.
Here $S_{P,th}$ is computed using the mean power tensor given in Eq. \ref{eq:meanAij}
and we subtract $\log 3$ since that is the value obtained in the limit of perfect
isotropy.

For $l=2, 3$ we find that, $\Delta S_P(2) = -0.0056$ and $\Delta S_P(3) = -0.064$
using the 9 year WMAP data. Similar results are obtained for 7 year data. For $l=2$
the observed value is very close to that expected in an isotropic model. For $l=3$
the deviation is larger but still less than $1\sigma$. Hence the dispersion in
eigenvalues for the observed data is not statistically significant for $l=2$ or
$l=3$. We nevertheless fix the value of anisotropy parameters in model II so as to
obtain a best fit to $\Delta S_P$ for $l=2,3$ by minimizing the square error.
In future it might be more appropriate to make a global fit to the CMBR data
in order to fix the parameters of this model.
The best fit is obtained for $\epsilon_H= -0.0054$, which leads to
$\Delta S_{P,th}=-0.034$ and root mean square error equal to 0.041. The value of $\Delta S_{P,th}$ is found to be independent
of $l$ since the anisotropy is equally effective for all the multipoles. The observed
data, however, shows dominant alignment only for $l=2,3$ multipoles. Hence we need
some mechanism to suppress the anisotropic contribution in the remaining multipoles.
This can be accomplished if the Universe quickly evolves into an isotropic de Sitter
phase during inflation \cite{Aluri2012b}. This mechanism is implemented in model III.

\begin{figure}[!t]
 \centering
 \includegraphics[scale=0.50,angle=-90]{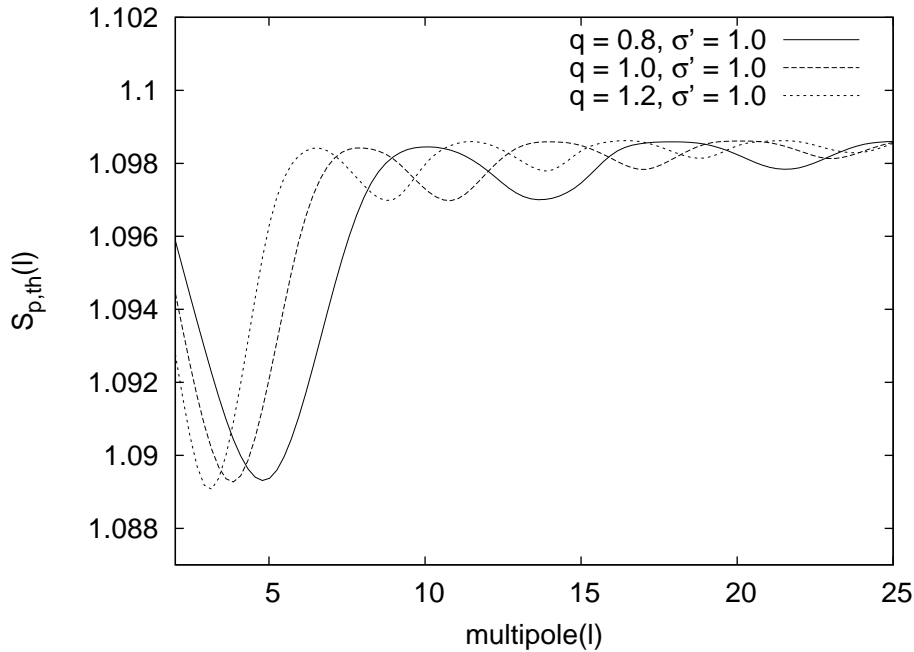}
 \caption{The power entropy as a function of $l$ for model III. The results
          are shown for three different choices of parameters, $\sigma'$ and
          $q=2/(\eta_0a_I\bar H)$.}
\label{fig:entropy}
\end{figure}

In model III, $\Delta S_{P,th}$ oscillates, with amplitude decreasing with $l$. We attempt to
fit the power entropy for $l=2,3$ in this model by fixing the parameter $q= 2/(\eta_0a_I\bar H)$
and determining $\sigma' = \sigma/a_{I}$ by minimizing square error. 
However we find that the best fit is driven to relatively large values
of $\sigma'$, where the perturbation theory becomes unreliable. The largest
perturbative contribution is obtained for the isotropic part of
the power spectrum. Hence here we only give results by fixing the parameter
$\sigma'=1.0$.
The results for $S_{P,th}$ as a function of $l$ for $q=0.8,\ 1.0,\ 1.2$ with $\sigma'=1.0$ 
are shown in Fig. \ref{fig:entropy}. The figure clearly shows the oscillations in $S_{P,th}$.
We also find that it approaches $\log 3$ for large $l$, as expected. 
The root mean square error for $\sigma'=1$ and  $q=0.8,\ 1.0,\ 1.2$ is found to be 
0.058, 0.056 and 0.055 respectively.  
We postpone a more detailed fit, involving larger values of $l$ and perhaps
polarization data in order to obtain a global minima to future research.
In Fig. \ref{fig:entropy} we notice that the anisotropic contribution becomes negligible
beyond $l\approx 10$. This might be related to the observations that low multipoles lying
in the range $2\le l\le 11$, and not just $l=2,3$, show alignment with the quadrupole
\cite{Copi2007,Samal2008}.

\section{Summary and Conclusions}
We have studied three models of anisotropic inflation in order to explain the observed
violation of isotropy in CMBR data. In models I and III, the Universe is anisotropic
only during the very early stages of inflation and quickly evolves into a de Sitter
space-time. In model II, the anisotropy is present throughout the period of inflationary
expansion. In all these models we choose the preferred direction same as the principal
axis of the CMBR $l=2$ mode. This axis points roughly in the direction of the Virgo
supercluster. Model I leads to direct correlations between multipoles $l$ and $l+1$.
Hence it leads to alignment between $l=2,3$. In models II and III, there is no direct
correlation between $l$ and $l+1$ modes. However each multipole is affected such that
its principal axis is aligned with the preferred axis of the model. Hence these models
also lead to alignment of the principle axis corresponding to $l=2, 3$ multipoles. In
models I and III, the anisotropic contribution affects dominantly the modes with low $l$.
For higher $l$ values, this contribution decays rapidly. Model II, in contrast, gives
equal contribution to all $l$ values.

We treat the anisotropic term perturbatively in all the three models. We compute the
power spectrum and measures of alignment of $l=2,3$ multipoles at first order in perturbation
theory. The measures of alignment use the concept of the power tensor \cite{Ralston2004,Samal2008}.
The three eigenvectors of this tensor define a frame in real space for each $l$. The eigenvector
corresponding to the largest eigenvalue defines the principal axis. In models II and III the
principal axis of both $l=2,3$ are aligned with the chosen preferred axis in these models
and hence are aligned with one another. In model II
we fix the theoretical parameters by
fitting the dispersion in eigenvalues in $l=2,3$ modes. In model I, the anisotropic term
does not contribute to the power tensor for any individual value of $l$. However it leads
to a correlation between $l$ and $l+1$ modes. We define a new statistic, which tests the
overall correlation of the power tensor between two adjacent modes. The theoretical estimate
of this statistic is fitted to the observed data for $l=2,3$ modes. We find that models I
and III provide a better description of 
data since they capture the important feature that the anisotropy
decays rapidly for large $l$. We also find that perturbation theory is qualitatively reliable in models I and II. 
 The contribution due to higher order terms in model I is likely to introduce a
correction of about 25\%. In model III, the best fit to the eigenvalue
dispersion for $l=2,3$ modes is driven to relatively
large values of the anisotropy parameter, making perturbation theory unreliable.
In any case this might not be the best method to fix the model parameters
since the dispersion observed in data is not statistically significant
for these modes. Hence it may be better to fix the parameters of
both models II and III by making a global fit to the CMBR temperature 
and polarization data. It may also be useful to perform a non-perturbative analysis
of the anisotropic metrics.

The observed data has shown several signals which indicate a preferred direction, pointing
roughly towards the Virgo supercluster. The fact that several diverse data sets, including
radio, CMBR and optical, lead to the same preferred direction gives us some confidence
that the observed anisotropy arises due to a physical effect. We have related these
observations to an early anisotropic phase of inflation. It would clearly be of interest
to test our models further by studying their implications for higher multipoles, CMBR
polarization, radio and optical polarizations and the large scale structure of Universe.

{\bf Acknowledgments:} We acknowledge the use of Legacy Archive for Microwave Background
data analysis. Some of the results of this work are derived using the publicly available
HEALPix package \cite{Gorski2005}.

\end{document}